\documentclass{mem}
\usepackage{natbib}\usepackage{txfonts}\usepackage{balance}
\usepackage{graphicx}
\usepackage[a4paper,breaklinks,dvipdfm]{hyperref}
\bibpunct{(}{)}{;}{a}{}{,}
\idline{75}{282}
\begin{document}
\def\teff{$T\rm_{eff }$}
\def\kms{$\mathrm {km s}^{-1}$}

\newcommand{\FeII}{[\ion{Fe}{ii}]}
\newcommand{\TiII}{[\ion{Ti}{ii}]}
\newcommand{\SII}{[\ion{S}{ii}]}
\newcommand{\OI}{[\ion{O}{i}]}
\newcommand{\OIp}{\ion{O}{i}}
\newcommand{\PII}{[\ion{P}{ii}]}
\newcommand{\NI}{[\ion{N}{i}]}
\newcommand{\NII}{[\ion{N}{ii}]}
\newcommand{\NIp}{\ion{N}{i}}
\newcommand{\NiII}{[\ion{Ni}{ii}]}
\newcommand{\CaIIp}{\ion{Ca}{ii}}
\newcommand{\PI}{[\ion{P}{i}]}
\newcommand{\CIp}{\ion{C}{i}}
\newcommand{\HeI}{\ion{He}{i}}
\newcommand{\MgIp}{\ion{Mg}{i}}
\newcommand{\MgIIp}{\ion{Mg}{ii}}
\newcommand{\NaI}{\ion{Na}{i}}
\newcommand{\HI}{\ion{H}{i}}
\newcommand{\brg}{Br\,$\gamma$}
\newcommand{\pab}{Pa\,$\beta$}

\newcommand{\macc}{$\dot{M}_{acc}$}
\newcommand{\lacc}{L$_{acc}$}
\newcommand{\lbol}{L$_{bol}$}
\newcommand{\mjet}{$\dot{M}_{jet}$}
\newcommand{\mh}{$\dot{M}_{H_2}$}
\newcommand{\Ne}{n$_e$}
\newcommand{\h}{H$_2$}
\newcommand{\um}{$\mu$m}
\newcommand{\lam}{$\lambda$}
\newcommand{\msyr}{M$_{\odot}$\,yr$^{-1}$}
\newcommand{\Av}{A$_V$}
\newcommand{\msun}{M$_{\odot}$}
\newcommand{\lsun}{L$_{\odot}$}
\newcommand{\cm}{cm$^{-3}$}
\newcommand{\ergscm}{erg\,s$^{-1}$\,cm$^{-2}$}

\newcommand{\bet}{$\beta$}
\newcommand{\alfa}{$\alpha$}

\hyphenation{mo-le-cu-lar pre-vious e-vi-den-ce di-ffe-rent pa-ra-me-ters ex-ten-ding a-vai-la-ble excited pro-duct no-ting}

\title{
Spatially resolved atomic and molecular emission from the very low-mass star IRS54}

   \subtitle{}

\author{
R. \,Garcia Lopez\inst{1} 
\and A. \, Caratti o Garatti\inst{1}
\and G. \,Weigelt\inst{1}
\and B. \,Nisini\inst{2}
\and S. Antoniucci\inst{2}
          }

  \offprints{rgarcia@mpifr-bonn.mpg.de}

\institute{
Max Planck Institut f\"{u}r Radioastronomie --
Auf dem H\"{u}gel, 69,
D-53121 Bonn, Germany
\email{rgarcia@mpifr-bonn.mpg.de}
\and
Istituto Nazionale di Astrofisica --
Osservatorio Astronomico di Roma, Via Frascati 33,
I-00040 Monte Porzio Catone, Italy
}

\authorrunning{Garcia Lopez }

\titlerunning{Spatially resolved H$_2$ emission}

\abstract{
Molecular outflows from very low-mass stars (VLMSs) and brown dwarfs (BDs) have been studied very little, and only a few objects have been
directly imaged.
Using VLT SINFONI K-band observations, we spatially resolved, for the first time, the \h\ emission around IRS54, a $\sim$0.1-0.2 M$_\odot$ Class I source. 
The molecular emission shows a complex structure delineating a large outflow cavity and an asymmetric molecular jet.
In addition, new \FeII\ VLT ISAAC observations at 1.644\,\um\ allowed us to discover the atomic jet counterpart which extends down to the central source.
The outflow structure is similar to those found in low-mass Class I young stellar objects (YSOs) and Classical TTauri stars (CTTSs). However, its \lacc/\lbol\ ratio is very high ($\sim$80\%), and the derived mass accretion rate is about one order of magnitude higher than in objects with similar mass, pointing to the young nature of the
investigated source.
\keywords{stars: formation -- stars: circumstellar matter -- ISM: jets and outflows -- ISM: individual objects: YLW52, ISO-Oph 182, IRS54 -- Infrared: ISM }
}
\maketitle{}

\section{Introduction}

\begin{figure*}[t!]
\resizebox{\hsize}{!}{
\includegraphics[clip=true]{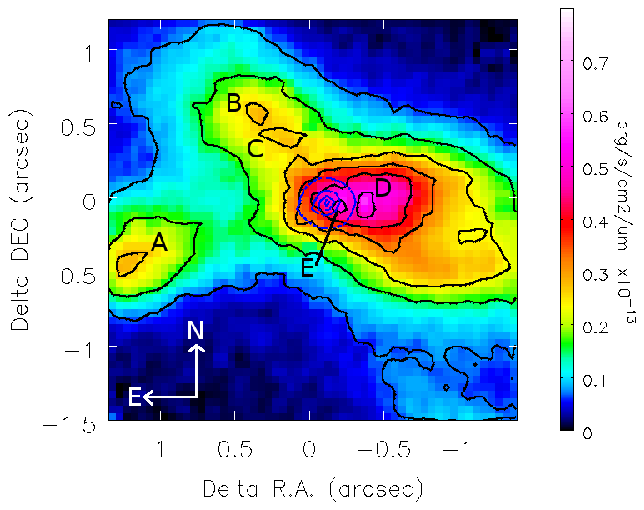}
\includegraphics[clip=true, totalheight=5cm]{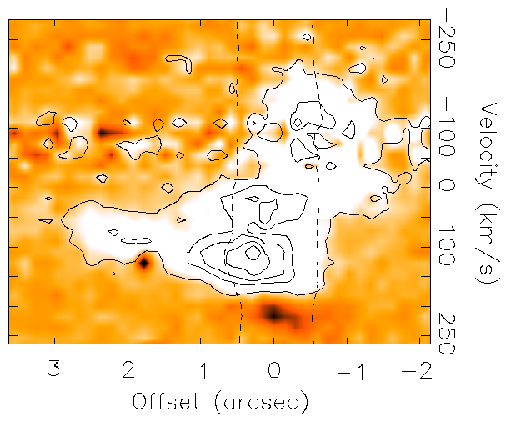}
}
\caption{\footnotesize
\textbf{Left: }Average emission over the H$_2$,1-0S(1) line in IRS54. Overplotted the positions of six condensations are indicated. For reference, contours of the continuum (near the H$_2$,2.122\,$\mu$m line) down to the FWHM size 
have been overplotted at the centre of the image (dashed-blue contours).
\textbf{Right:} Continuum-subtracted position velocity diagram of the \FeII\,1.644\,$\mu$m along IRS54. The slit was centred on-source encompassing positions E and D (i.e., with position angle $\sim$ 90$^o$ ). Radial velocities are measured with respect to the local standard of rest (LSR) and corrected for an average cloud velocity of 3.5\,\kms\ \citep{wouterloot05,andre07}. For reference, contours of the continuum were overplotted with dashed lines.}
\label{all_chanels}
\end{figure*}

Very low-mass stars and brown dwarfs are thought to possess circumstellar disks when they are born, independently of the formation mechanism \citep{luhman12}. Outflows and protostellar jets are then expected in these sources as an outcome of the accretion process.   

The first jet from a BD was detected in 2005 \citep{whelan05}, and since then, a few objects have been studied. 
Most of them have been investigated through spectro-astrometry of forbidden emission lines \citep[FELs][]{joergens12,whelan09}, and only recently, through direct CO imaging \citep{phan11,phan08}.
These studies mostly involve  relatively evolved TTauri-like objects, whereas young embedded sources (the so-called Class 0 and I objects) have seldom been  studied mainly due to the lack of young candidates.


Jets from embedded young sources were traditionally identified 
through H$_2$ observations at 2.122\,\um.  
Although large-scale protostellar jets are not expected in VLMSs or BDs, molecular hydrogen emission line regions (MHELs) could be present around these sources, in analogy to low-mass Class I objects \citep{davis_MHEL}. 

We present here SINFONI K-band integral-field spectroscopic observations of the Class I VLMS IRS54 (YLW\,52) at medium resolution. This source ($\alpha$=16:27:51.7, $\delta$=-24:31:46.0) is located outside the main clouds in the Ophiuchus star-forming region, and it has  a bolometric luminosity of only $\sim$0.78\,\lsun\ \citep{vankempen09}. In addition, IRS54 is a strong candidate to drive an outflow and a  jet as revealed by Spitzer and H$_2$ narrow-band images around the source \citep{tigran04,jorgensen09}.  


\begin{figure*}[t!]
\resizebox{\hsize}{!}{
\includegraphics[clip=true]{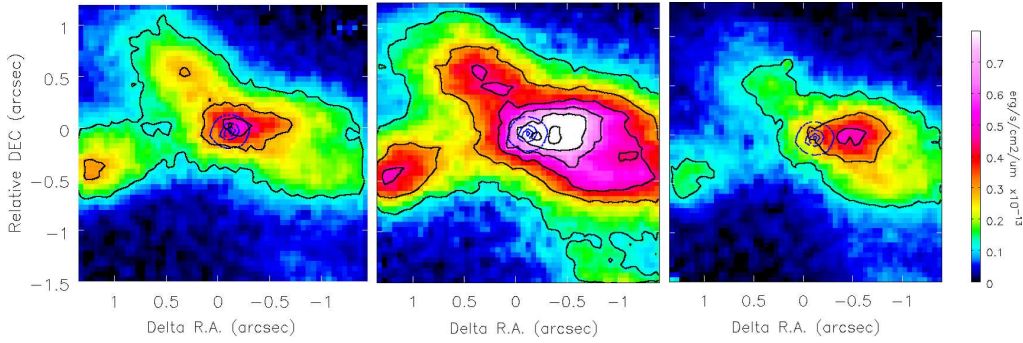}
}
\caption{\footnotesize 
Continuum-subtracted H$_2$\,1-0S(1) images of IRS54. From left to right: 
average over two spectral channels at $-64$\,\kms\ and -30\,\kms\ (left), and +40\,\kms\ and +75\,\kms\ (right). A single spectral channel at +6\,\kms\ is plotted in the centre. 
The velocities are with respect to the LSR and corrected for the cloud velocity. }
\label{velocity_chanels}
\end{figure*}

\section{Results}
\subsection{H$_2$ outflow morphology}

\begin{figure}[t!]
\resizebox{\hsize}{!}{
\includegraphics[clip=true]{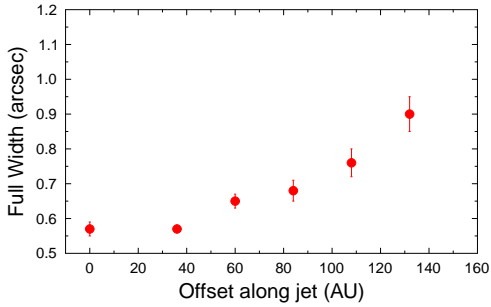}
}
\caption{\footnotesize Width of the \h\,1-0S(1) jet-like structure as a function of the distance from the source.The jet width has been estimated fitting a single Gaussian function to vertical cuts across the H$_2$ emission.}
\label{opening_angle}
\end{figure}

Figure\,\ref{all_chanels} (left panel) shows the \h\,1-0S(1) continuum-subtracted spectral image of IRS54 integrated across five velocity chanels. The morphology is very complex with gas displaying an X-shaped spatial distribution superimposed on a more collimated structure, possibly jet-like (see discussion below), located westward of IRS54 and extending over $\sim$1\arcsec\ ($\sim$120\,AU). All regions show condensations at positions A, B, C, D and E, displaced from the source ($\Delta \alpha$,$\Delta \delta$) at about (1\farcs3,-0\farcs4), (0\farcs4, 0\farcs5), (0\farcs3, 0\farcs4), (-0\farcs1,0), and (-0\farcs3,0), respectively.
As shown in the channel maps in Fig.\,\ref{velocity_chanels}, there is no clear blue- or red-shifted lobe. Indeed,  red- and blue-shifted emission can be associated roughly with the same spatial regions. 
The \h\,2.122\,\um\ line shows a very broad profile with a full width zero intensity of $\sim$200\,\kms, but with a peak velocity around $\sim$0\,\kms. This, along with the intrinsic \h\ low velocity, may indicate that the outflow is very close to the plane of the sky.
Figure\,\ref{velocity_chanels} clearly shows that the jet-like structure is only present in one lobe of the outflow. This might indicate different excitation conditions for the two lobes, i.e. an asymmetric jet in which the eastern lobe (likely the blue-shifted one; see Sect.\,2.2 and the right panel of Fig.\,\ref{all_chanels}) might have a higher velocity, dissociating the molecular \h\ component. 
On the contrary, it might be that the jet-like structure is just H$_2$ gas excited along the outflow cavity. However, the measured width of the  \h\,1-0S(1) jet-like structure as a function of the distance from the source is consistent with a full opening angle of the flow of only $\sim$23\,\degr\ (see Fig.\,\ref{opening_angle}). This value is very similar to those found in jets from low-mass Class I sources and CTTSs, which show opening angles between 20\degr\ and 42\degr\ \citep{hartigan04,davis11}

\subsection{The atomic jet component}

The hypothesis of an asymmetric jet in the eastern lobe able to dissociate the molecular \h\ component can be tested by detecting ionic/atomic emission eastwards of IRS54. With this in mind, we asked for new VLT-ISAAC H-band observations at medium resolution. The slit was aligned along the \h\ jet-like structure, i.e., PA=90\degr\ on the sky.
The right panel of Fig.\,\ref{all_chanels} shows the position-velocity diagram (PVD) of the \FeII\,1.644\,\um. The YSO continuum was removed in order to study the jet structure close to the driving source. As shown in the figure, a blue-shifted atomic jet is detected eastward of the source. In addition, the PVD shows the presence of two velocity components close to the source, the so-called high- and low-velocity components (HVC, LVC), with v$_r^{high}\sim$-125\,\kms\ and v$_r^{low}\sim$-25\,\kms. 
Analogous to low-mass Class I sources and CTTSs, the LVC is confined around the source position, while the HVC extends further away from the driving source \citep{rebeca08, rebeca10}.

\section{Accretion and ejection properties}

From the \brg\ emission line detected in our SINFONI spectra, it is possible to derive the accretion luminosity (\lacc) and mass accretion rate (\macc).
Using the relation between the luminosity of the \brg\ line (L(\brg)) and the accretion luminosity (\lacc) derived by \cite{calvet04}, we found an accretion luminosity of \lacc$\sim$0.64\,\lsun. The luminosity of the \brg\ line was measured from the integrated flux across the \brg\ line in our cube (F=4.08$\times$10$^{-14}$\,\ergscm) and corrected by a visual extinction of 30\,mag. 
The accretion luminosity towards the source is very high, especially when compared with its bolometric luminosity of $\sim$0.78\,\lsun. This results in a \lacc/\lbol\ value of $\sim$80\%, consistent with a very young and active YSO. 

\begin{figure}[t!]
\resizebox{\hsize}{!}{
\includegraphics[clip=true]{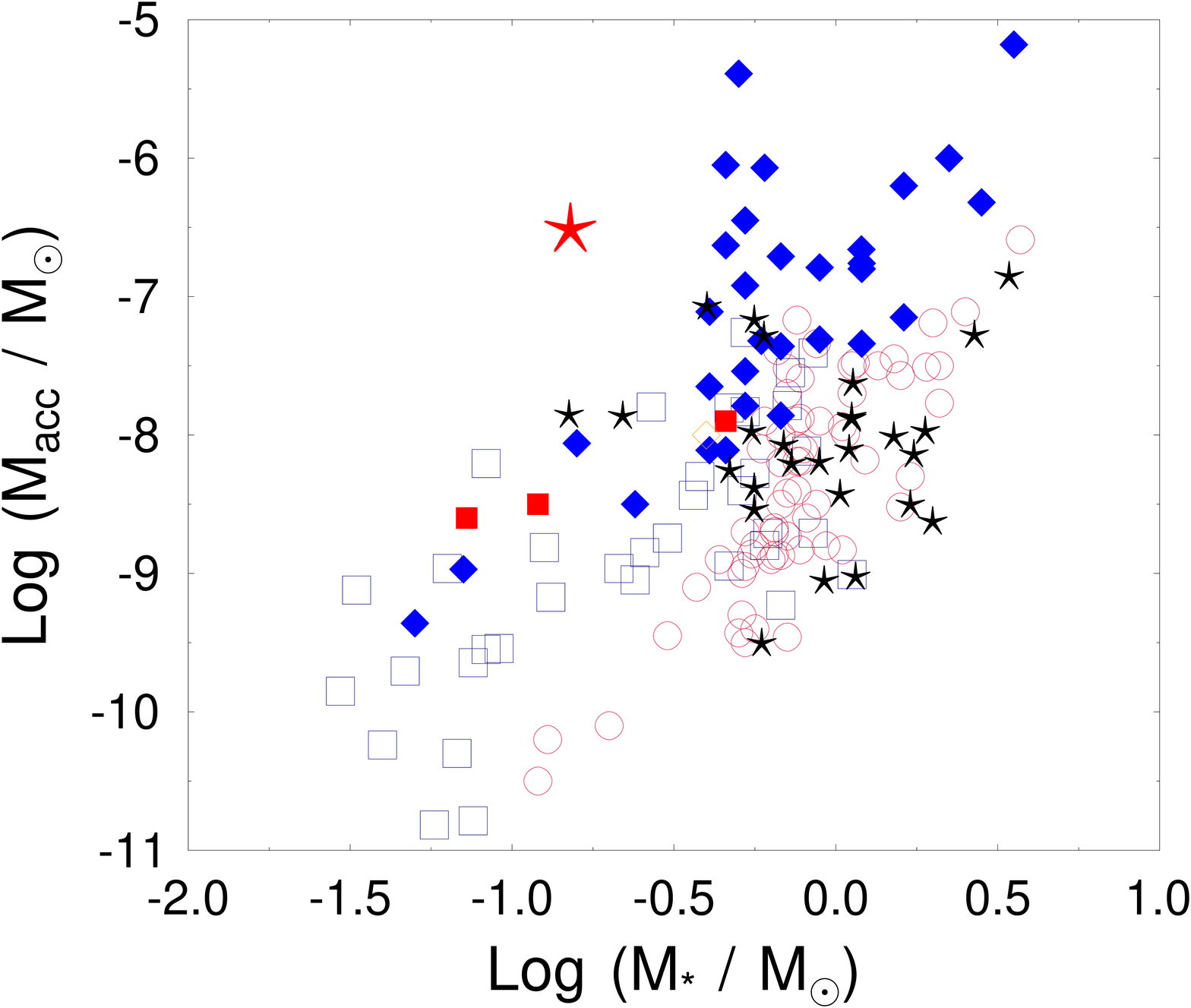}
}
\caption{\footnotesize Mass accretion rate as a function of M$_*$ for IRS54 (big red star); Ophicus, Taurus-auriga and L1641 Class II objects (open squares, open circles and stars; \citealt{natta06,white01, calvet04, aleL1641}); very low-mass stars and brown dwarfs in Taurus-auriga (filled squares; \citealt{white03}); and Class I sources (filled diamonds; \citealt{white04,simone08}) }
\label{macc}
\end{figure}

From the computed \lacc\ value, a mass accretion rate of $\sim$3.0$\times$10$^{-7}$\,\msyr\ is inferred, using  \macc=(L$_{acc}$\,R$_*$\,/\,G\,M$_*$)$\times$(1-R$_*$/R$_i$)$^{-1}$ \citep[with R$_i$ =5\,R$_*$, see][]{gullbring98} and the stellar parameters from \cite{rebeca13}.
Figure\,\ref{macc} shows \macc\ (IRS54: big red star) as a function of the stellar mass. For comparison, the values derived by several authors for a large sample of Class II and I objects  are also represented.   
The \macc\ value of IRS54 is higher than those found in objects of roughly the same mass, pointing again to the young nature of this source.

In order to compare the \macc\ value of IRS54  with the mass transported by the jet, we have computed the mass loss rate transported by the molecular (\h) and atomic (\FeII) components.
In order to derive the mass loss rate, the tangential velocity of the outflow must be known. Because the inclination angle of the jet is unknown and the small radial velocities suggest that the outflow is almost on the plane of the sky, we assume a lower limit to the jet velocity of $\sim$200\,\kms. This value corresponds to the FWZI of the \h\ and \FeII\ lines.
The mass transported by the warm molecular component was then computed using an average H$_2$ column density of N(\h)$\sim$1.7$\times$10$^{17}$\,cm$^{-2}$ \citep{rebeca13} and following the expression \mh=2\,$\mu$\,m$_H$N(H$_2$)\,A\,dv$_t$/dl$_t$ (see, e.g., \citealt{davis_MHEL,davis11}). Here, A is the area of the emitting region, $\mu$ is the mean atomic weight, and dl$_t$ and dv$_t$ are the projected length and the tangential velocity. Finally, we have assumed A equal to the extent of the flow along the jet axis from position S to D ($\sim$0\farcs5) multiplied by the width of the flow (i.e. the seeing).
These assumptions result in a value of \mh$\gtrsim$1.6$\times$10$^{-10}$\,\msyr. 

Following similar considerations and employing the expression $\dot{M}_{[FeII]}$=$\mu$\,m$_H\times$(n$_H$ V)$\times$v$_t$/l$_t$  \citep[see][]{nisini_HH1}, we have derived the mass transported by the atomic component. In addition to the parameters defined above, n$_H$ and V are the total density and the volume of the emitting region. The product of both quantities can be expressed as (n$_H$ V)= $L(line)(h \nu A_i f_i \frac{Fe^+}{Fe} \frac{[Fe]}{H})^{-1}$, where $A_i$ and $f_i$ are the radiative rates and fractional population of the upper level of the considered transition and $Fe^+/Fe$ is the ionisation fraction of the iron having a total abundance with respect to hydrogen of $[Fe/H]$.
To compute the fractional population, we have used an electron density value n$_e$=18\,000$^{+17\,000}_{-8500}$\,\cm\ derived from the \FeII\,1.600/1.644\,\um\ line ratio \citep[see, e.g.,][for more details on the employed method]{rebeca08}.  
We have assumed that all the iron is ionised, and a total abundance equal to the solar one \citep{asplund05}. This is equivalent to consider no dust depletion and thus yields to a lower limit of $\dot{M}_{[FeII]}\gtrsim$2.0$\times$10$^{-8}$\,\msyr.
The derived $\dot{M}_{[FeII]}$ value is roughly two orders of magnitude higher than that computed for the warm molecular component. Similar results have been also found in low-mass Class I sources \citep[e.g.,][]{davis11}, probably indicating that the atomic component is transporting most of the mass ejected by these sources.
The computed \mh\ and  $\dot{M}_{[FeII]}$ values are from two to three orders of magnitude lower than the \mh\ and $\dot{M}_{[FeII]}$ values derived for low-mass Class I sources using the same technique \citep{davis11}. However, it is worth noting that those sources are more massive than IRS54. 

Finally, the $\dot{M}_{[FeII]}$ and \macc\ ratio is $\sim$0.1, in agreement with the predictions of MHD wind models. 


\section{Conclusions}

We presented the first spatially resolved \h\ (MHEL) and \FeII\ (FEL) emission regions around IRS54, a Class I VLMS. Both the atomic and molecular emissions were detected down to the central source (within the first $\sim$50\,AU).
The \h\ emission might be interpreted as coming from the interaction of a wide-angle wind with an outflow cavity and a molecular jet, while the \FeII\ emission is clearly tracing an atomic blue-shifted jet eastward of IRS54.
The derived \lacc\ and \macc\ values for this source are high, especially when compared with sources of roughly the same mass. This strongly suggests that IRS54 is a young VLMS still accreting mass at a high rate. 
Although the mass loss rate computed for IRS54 is lower than in low-mass Class I sources, the outflow/jet morphology is very similar. In addition, we found that the $\dot{M}_{[FeII]}$/\macc\ ratio is roughly 0.1, in agreement with MHD wind model predictions. 
Our results might then indicate that there is a smooth transition from low-mass to VLM jets and outflows.

\bibliographystyle{aa}
\bibliography{references}

\end{document}